\documentstyle[12pt]{article}

\newcommand{\s}{\sigma }
\newcommand{\ve}{\varepsilon }

\newcommand{\Ll}{\Lambda}
\newcommand{\be}{\begin{eqnarray} }
\newcommand{\ee}{\end{eqnarray} }

\begin{document}
\baselineskip 14pt

\begin{center}
{\Large \bf
Critical properties of Toom cellular automata
\bigskip

Danuta Makowiec {\it fizdm@univ.gda.pl}
}

\end{center}
\abstract{
The following paper is the continuation of  our earlier considerations on  
cellular automata with Toom local rule (TCA)  as the alternative to 
kinetic Ising systems.  The arguments  
for TCA stationary states not 
being the equilibrium states are found in simulations. 
}

\section{Motivation}
The Monte Carlo method applied to statistical physics denotes
simulation of some stochastic system for which time averages restore 
equilibrium ensemble expectations\cite{Binder,Stauffer,Binney}. 
Although the dynamical system obtained in this way  is often completely 
artificial  to the original equilibrium system, but  much freedom 
in  designing dynamics  offers possibility to verify different 
hypothesis about micro scale interactions in the system. In this way, 
the Monte Carlo simulations allows  to examine links between the  
micro dynamics and resulting equilibrium system.
In case of the widely known  Lenz model of interacting spins, the 
Monte Carlo method provides the so-called kinetic Ising models.

Why not search among other dynamical systems,
such systems which also can mimic properties of some equilibrium 
system ?

The proposition originated from cellular automata is qualitatively 
distinct from the mentioned kinetic models. The cellular automata 
 are complex  dynamical systems. It means
that we are given the set of local rules instead of the  
notion of energy and changes in the system, means evolution, are 
synchronised. The steps of synchronised update are interpreted 
as time steps. Therefore, the main goal in study such systems is to 
give meaning to the standard thermodynamic notions like energy, 
pressure, specific heat, temperature, {\it etc.} 
\cite{BG,LMS,Lebowitz}. 
In the case when the local rule is not reversible this goal is not 
obvious 
\cite{LMS,MaesKoen}. Especially, little is known about the nature 
of the stationary measures in the regime where there is more than 
one stationary measure.

In the following we continue our study of  cellular automata with 
Toom local rule (TCA)  as the alternative to kinetic Ising systems 
\cite{PRE,Zak97}.  
This model is known to exhibit non-ergodic properties for certain model 
parameters \cite{Toom,BG,LMS,MaesKoen}. Therefore it can mimic the 
system undergoing the continuous phase transition. 

\section{ Toom cellular automata}
For every spin $\s_i \in \{ -1, +1 \} $ attached to the $i$ node
of the  
square lattice $Z^2$ we choose its three nearest-neighbours, named 
$N_i, E_i, C_i$ as follows:
$$\begin{array}{cccccccc}
  &|&  &|&  &|& \\
- &. &-&{N_i} &-&. &-&.\\
 &|  && | && | && \\
- &. &-&{C_i=\s_i } &-&{E_i} &-&.\\
 &|  && | && | && \\
\end{array}$$

They vote totally for  the $i$ spin state in the next time step,
namely:

Let $\Sigma_i = {N_i} +{E_i} +C_{i} $
then
\be
 \s_i(t+1)= \biggl\{ \begin{array}{cc}
{\rm sign}\ \Sigma_i &{\rm with\ probability} \quad {1\over 2}(1+\ve)\\
\\
- {\rm sign}\ \Sigma_i &{\rm with\ probability} \quad {1\over 2}(1-\ve)\\
\end{array} 
\label{toom}
\ee
The parameter $\ve \in [0,1] $  mimics the stochastic temperature 
effects:
$\ve=1$ means completely deterministic evolution, $\ve=0$
corresponds to the random rule .

One may wish to compare Toom dynamics to the Domany rule--- 
cellular automata with this rule  provide the equilibrium system 
\cite{LMS}:
\be
\s_i(t+1)= \Biggl\{ \begin{array}{cc}
{\rm sign}\  \Sigma_i & {\rm with \ probability } \quad 
\biggl\{ \begin{array} {cc} 
1-\ve_1^D & {\rm for} |\Sigma_i|=1\\
& \\
1-\ve_3^D & {\rm for} |\Sigma_i|=3\\
\end{array} \\ 
-{\rm sign}\  \Sigma_i & {\rm with \ probability } \quad 
\biggl\{ \begin{array} {cc} 
\ve_1^D & {\rm for} |\Sigma_i|=1\\
&\\
\ve_3^D & {\rm for} |\Sigma_i|=3\\
\end{array} \\
\end{array} 
\label{domany}
\ee
and $\ve$ the temperature-like parameter provides 
$$ \ve_1^D={1\over 2} (1-{\rm tanh}\ \ve) \qquad {\rm and } \qquad 
\ve_3^D={1\over 2} (1-{\rm tanh}\ 3\ve) $$

One can notice two "temperatures" playing in the Domany model. The 
lower temperature $\ve_3^D$ is assigned to the homogeneous with 
respect to the spin state areas what protects them,  while 
the higher temperature acts "kicking" more frequently the mixed 
neighbourhoods.

\section{Critical properties of TCA}
The renormalization procedure applied to equilibrium statistical 
mechanics systems provides the scaling laws, i.e. gives the 
relations satisfied by the critical exponents $\alpha, \beta, 
\gamma, \delta, \nu, \eta $ \cite{Binney}. Each critical exponent 
characterises the power law dependence of some observable when the 
system is undergoing a continuous phase transition.
Applying the standard methods for estimate singularities 
\cite{Klamut,Landau} we have  found the following four 
of the critical exponents:
\begin{itemize}
\item[$\beta $]--- magnetisation exponent:
 \qquad $  m(\ve ) \sim  (\ve- \ve_{cr})^{\beta} 
\quad {\rm at} \quad \ve > \ve_{cr} $
\item[$\gamma $]--- magnetic susceptibility exponent:
 \qquad $\chi(\ve)  \sim  |\ve - \ve_{cr} |^{-\gamma} 
\quad {\rm at} \quad \ve \rightarrow \ve_{cr}  $
\item[$\nu$]--- correlation length exponent 
 :\qquad $\xi(\ve) \sim |\ve-\ve_{cr}|^{-\nu} $ at 
$\ve\rightarrow \ve_{cr}$
\item[$\eta$]--- correlation decay at the critical point:
 \qquad $C(i,j) \sim |i-j|^{-\eta } $ at  $\ve=\ve_{cr}$         
\end{itemize}

Our results and the methods used to obtain then are presented in Figs.1-4. 

It is easy to check that with our estimations, namely: \\
\begin{itemize}
\item $\ve_{cr}= 0.822 \pm 0.002$
\item $\beta = 0. 12\pm 0.03$
\item $\gamma = 1.75 \pm 0.03$
\item $\nu= 0.88 \pm 0.02$ 
 \item $\eta=0.56\pm 0.02$
 \end{itemize}
the scaling lows:
\begin{itemize}
\item
$ \gamma =\nu(2-\eta)$ 
\item $ \beta= {1\over 2} \nu \eta $
\end{itemize} are not 
satisfied. 

One can notice that our values obtained for $\beta$ and 
$\gamma$ are in the good agreement with those characterising the 
two-dimensional Ising system while the rest critical parameters $\nu 
$ and $\eta$ are much different from the corresponding ones in the Ising 
system. 

Hence, {\bf TCA system seems to  neither belong to the Ising 
class of universality nor be a equilibrium system }.

\section{Features of TCA measures}

Statistical mechanics offers tools to investigate if a given 
stationary state can be represented by some
 Gibbs measure \cite{Enter,MaesKoen,MaesKoenPreprint}.
 For a probability measure $\mu$ to be a Gibbs 
measure denotes that for any finite configuration $\{ \s_\Ll \} $, 
$\quad \ln \mu( \s_\Ll ) $ exists and means the energy carried by the 
configuration $\{ \s_\Ll \} $. 
The two basic features of Gibbs measures are the, so-called, 
quasilocality of interactions and the proper properties of large 
deviations.

{\bf A: Quasilocality }

The idea of the quasilocality is shown in Fig.5. Using the 
notation used in Fig.5 one can say that nothing can arrive to some 
finite area $\Ll$ from infinity without changing the $\s_\Gamma$  
configuration. In particular, if a measure $\mu$ is quasilocal then 
for the average magnetisation one have
\be
| < m(\{ \s_\Ll \} _{|\s_\Gamma}) >_\mu  -
< m(\{ \s'_\Ll \} _{|\s_\Gamma}) >_\mu  | 
\longrightarrow_{ \Gamma \rightarrow \infty} 0 .
\label{quasi}
\ee

In the following experiments we test the presence of the above 
property in TCA.

\begin{itemize}

\item[{\bf I}] Let $\s_\Gamma $ be the fixed configuration evolving
according to the Toom dynamics at some $\ve$. Let $\s_{\Gamma^c}$ 
--- the boundary configuration,  be :

\item[a)] the homogeneous configuration of all spins $+1$

$$
\begin{array}{ccccccc}
+&+&+&+&+&+&\\
+&. & m_{-}^\ve  &. &. &+&\\
+&. &. &\bf O &. &+&\\
+&. & m_{+}^\ve  &. &. &+&\\
+&+&+&+&+&+&\\
\end{array} {{\rm TCA} \atop \longrightarrow }
\begin{array}{cccccccc}
+&+&+&+&+&+&\\
+&. &. &. &. &+&\\+&. &m_{+}^{\ve+\ve'}&\bf O &. &+&\\
+&. &. &. &. &+&\\
+&+&+&+ &+&+&\\
\end{array} $$
Then the $(+)$ boundary built from the rectangle of pluses {\it rises} 
the probability to find a spin in $+1$ state at the origin $\Ll=\bf 
O$ and $\ve'$ 
depends on the distance from the right and top sides of the 
rectangle (see Fig.6 for details).\\
{\it Remark:} If $\Gamma$ is not of the rectangle shape then the Toom 
interactions enlarge $\Gamma$ to have the right angle between the 
bottom and left sides.

\item[b)] the flat-interface configuration:  $+1$ on the right and 
$-1$ on the left of the flat-interface:

$$
\begin{array}{cccccccc}
-&-&-&+&+&+&\\
-&. & m_{-}^\ve  &. &. &+&\\
-&. &. &\bf O &. &+&\\
-&. &m_{+}^\ve  &. &. &+&\\
-&-&-&+ &+ &+&\\
\end{array} {{\rm TCA} \atop \longrightarrow }
\begin{array}{ccccccccc}
-&+&+&+&+&+&\\
-&. &. &. &. &+&\\-&. &m_{+}^{\ve +\ve'} &\bf O &. &+&
\\-&. &. &. &. &+&\\
-&+&+&+ &+&+&\\
\end{array} $$

Then the position of the flat-interface moves  to the left side of 
the rectangle and stays there thanks to the periodic boundary 
conditions. The dependence $\ve'$ on lattice site is the same as in 
the previous experiment.
\item[{\bf II}] Let $\s_{\Gamma^c} $ evolves along TCA rule at some 
$\ve_{out} >0$. Let the initial state for both  $ in$- 
and $out $- configurations is $(+)$. Let $\ve_{out} $ be chosen such 
that it generates the stationary state dominated by $(+)$ phase, 
$\ve_{out} \gg \ve_{cr} $. Then we observe:

$$ \begin{array}{lcccl}
in-{\rm state} & & out- {\rm state}& & {\rm resulting \ state} \\
\ve_{in} \gg \ve_{cr} &  &
&&m_\Gamma(l)=m(\s'{\rm \ typical\ for \ } \mu_+^{\ve_{in} }),\ \\
&&&&m_{\Gamma^c}(l)=m(\s'{\rm \ typical\ for \ } \mu_+^{\ve_{out} }) \\
\ve_{in} \approx \ve_{cr} & & 
&&m_\Gamma(l)=m(\s'{\rm \ typical\ for \ } \mu_+^{\ve_{cr}+\ve' }),\ \\
&&\ve_{out} \gg \ve_{cr} & 
& m_{\Gamma^c}(l)=m(\s'{\rm \ typical\ for \ } \mu_+^{\ve_{out} }) \\
\ve_{in} \approx \ve_{cr} +\ve' & & &  
& m_\Gamma(l)=m(\s'{\rm \ typical\ for \ } \mu_+^{\ve_{cr} }) = 0,\ \\
&  &\Longrightarrow & &m_{\Gamma^c}(l) \neq 0 \quad {\rm RANDOM} \\
\ve_{in} \ll \ve_{cr} & & & 
& m_\Gamma(l)= m_{\Gamma^c}(l)=0  \\
\end{array}$$

These results together with the case when the initial $out $- phase 
is $(-)$ are presented in Fig.7. Let us comment our observation as 
follow:
\begin{itemize}
\item After adjusting the phase of the $in$ configuration, the 
regions of distinct temperatures evolves independently. The 
two-point correlation function dies in one lattice unit.
\item The $out$-phase picks up the $(+)$ phase from the $in$-system 
when the $in$-system  is in the phase transition regime.
\item The large homogeneous structures created in the $in$-system 
undergoing  the phase transition, propagates freely to the $out$ 
configuration. Since the phase of these structures is random, then 
the phase of the $out$ configuration  changes randomly.
\item The random $in$-system  propagates random errors outside 
destroying the $out$- phase magnetisation.
\end{itemize}

{\it Remark:} The case when $\ve_{out} \ge \ve_{cr} $ provides always 
$ m_\Gamma(l)= m_{\Gamma^c}(l)=0 $
\end{itemize}

{\bf Concluding:}
\begin{itemize}
\item If $\ve_{in} < 0.76 $ then the conditional distribution for the
magnetisation in $\Gamma$ is independent of $\Gamma$- configuration.\\
Instead, there is observed dependence in $\sigma_{\Gamma^c}$  on  random 
clusters created in $\s_\Ll$ if $\ve_{in}$ is about the limit value and
$\ve_{out} \gg \ve_{cr} $.
\item If $\ve_{in} >0.76$ then the conditional distribution  for the
magnetisation depends on the system outside, in the sense that the 
probability for  the spin at the origin $o$ to take $+1$ state 
satisfies the inequality:
\be 
 \vline\quad
 {\rm Prob}\{ \s_O=+1 &|\ \s_\Gamma(\ve_{in})\ {\rm and}\ 
\s_{\Gamma^c}(\ve_{out}') \} \nonumber\\
- {\rm Prob} \{ \s_O=+1\  &|\  \s_\Gamma(\ve_{in})\ {\rm and}\  
\s_{\Gamma^c}(\ve_{out}'') \} \quad\vline >0 
\nonumber
\ee
So that the conditional probabilities are discontinuous, what in the 
lattice system topology denotes that the property (\ref{quasi}) is 
not satisfied. The stationary measure  might be  strongly 
non-Gibbsian \cite{MaesKoenPreprint}. However, if $\ve_{in} >0.84 $ 
then after 
thermalization time, what means  time allowing system for  adjusting 
the phase of the $in$-configuration, both areas evolve independently 
of each other.
\end{itemize}

{\bf B. Large deviations}

For stationary measures arisen from any Markov process to be or not 
to be the Gibbsian ones is determined by the zero or non-zero value of 
the relative entropy density $i(\mu|\nu) $ between different 
stationary  measures $\mu$ and $\nu$ of the system considered.
If $i(\mu|\nu) > 0 $ then both measures are  non-Gibbsian 
\cite{Enter,MaesKoen}.  
Thanks to the large deviation theorems  \cite{Enter} we have 
the  powerful way to estimate $i(\mu |\nu )$ .

In case of
$\mu_- $ and $\mu_+ $:  two stationary TCA measures corresponding to 
$(-)$ and $(+)$ phases, respectively, 
$i(\mu_- |\mu_+ ) $  can be  extracted from  the 
probability of  the  large fluctuation event, namely, from the 
probability that the large area of spins with negative 
magnetisation  occurs in the stationary state described by the 
$\mu_+$ measure.

From computer experiments we collect data on the magnetisation of 
square blocks of the size  $l\times l$.
 Then on the base of the formula
 \be
i_l(\mu_- |\mu_+ ) =
\lim_{l\rightarrow\infty} 
{1\over l^2} 
\ln {\rm Prob}_{\mu_+} \{ m(\s_{l\times l}) <0 \} 
\ee
we try to estimate the limit
$$
i(\mu_- |\mu_+ ) =
\lim_{l\rightarrow\infty} i_l(\mu_- |\mu_+ ) .$$

Fig.8 is to present the density of the relative 
entropy $i_l(\mu_- |\mu_+ ) $ for different   lattice sizes: 
$L= 60, \ 100,\ 200$. 
Although averaging time was very large (10 000 time steps), the 
strong time dependence is noticeable. This dependence moves to the 
calculations of $i_l(\mu_- |\mu_+ ) $ causing difficulties in 
estimating the limit $ l\rightarrow \infty$. 
Anyway, we tried to find out the  block size where the relative 
entropy density  arrives at the zero.
 According to the presented results for the lattice size $L=200$
 this limit  should occur if the blocks 
 are of the size $l > 50 $.
In the next figure, Fig.9,  we present  the failure of 
this suggestion.  What we observe in simulations is 
that with the increase of 
the block size the relative entropy density decay slows down.

Moreover, we test the density of relative entropy between the 
stationary measure arising in TCA evolving  little apart from the 
critical point, namely at $\ve=0.800$ and the $(-)$ minus phase. 
(see Fig.10 for these results). It appears that  at $l=105$
 the relative entropy density  would reach the zero.
But this block  size   is greater than the lattice size. 

Concluding, we can say that {\bf the relative entropy density
between stationary measures of Toom cellular automata evolving in
the critical regime on the periodic lattice is positive.}

\section{Conclusions}
Although the critical regime in TCA manifests itself in the way 
characteristic to any thermodynamic system, i.e. by the rapid 
increase in the two-point-correlation function  of spin states, but 
the phenomena driving  the system  seems to be 
different from any equilibrium system.

The TCA system  undergoing the phase transition stays in the
extremely dynamically fragile state. There is kept the sensitive 
balance between two processes: the processes of  self-organising 
spins state to enlarge pure phase clusters and the stochastic 
process which destroys this clusters. 
Since the homogeneous clusters are not specially protected as it 
happens in the Domany CA, the area of cluster of one phase varies. 
In consequence, the correlations between spins are damped much 
stronger than in Domany CA and Ising kinetic models. However,  it is 
astonishing that the relations between spin sites do not influence 
the critical behaviour of  the order parameter.
Therefore, one can say that the phenomenological conjecture that all 
two-dimensional ferromagnetic systems belongs to the same 
universality class --- the  class of Ising system \cite{Grinstein}, 
is  "partially" satisfied ( if the participation into the 
universality class could be partial). 

Testing locality of interactions we  have found that at about the 
critical point there exists a boundary for the interaction to be 
finite. If the stochastic perturbation is sufficiently strong than 
the information from distant spins is lost. Otherwise, no matter how 
far away spins are, their influence on each other is evident.

Our examinations on large deviation properties as well as 
quasilocality have hit the boundary of the finite lattice size.  
Therefore one can say that the system we studied does not restore 
properties of any infinite system. Such a conclusion often 
accompanies complex system considerations. Complex systems, existing 
on the, so-called, {\it edge of order and chaos} \cite{edge}, are 
known for that nobody is able to predict what kind of phenomena will 
arise in a system if one moves a little the system parameters.
Studying critical properties of Toom cellular automata we obtain 
arguments for the suspicion that self-organised criticality coming 
from the dynamic equilibrium provides systems qualitatively distinct 
from the thermodynamic systems \cite{complex}.

\newpage
\begin{figure}
\caption{ Estimates for $\beta $ in Toom CA. 
Data collected in 430, 280, 120 experiments with lattices: $L=100,\ 
150,\ 200$ respectively.
}\end{figure}

\begin{figure}
\caption{ Estimates for $\gamma$ in TCA from the relation 
$\kappa = <m^2> -<m>^2 $. Data collected 
in 430, 180, 120 experiments with lattices: $L=100,\ 120,\ 150 $ 
respectively.
}\end{figure}

\begin{figure}
\caption{ Estimates for $1\slash\nu$ in TCA on the base of finite size 
theory from the maximal slope of derivative of the 4th order 
magnetisation cumulant as well as from the maximum slope of the 
logarithm derivatives of: absolute value of magnetisation, square 
magnetisation, fourth power of magnetisation.
Data collected in 1400, 860, 660, 430, 280, 120 experiments with 
lattices: $L=20,\ 30,\ 60,\  100,\ 150,\ 200 $ respectively.
}\end{figure}

\begin{figure}
\caption{The decay rate $\eta$ for the two-point correlation 
function of magnetisation obtained on the lattice with 
$L=100$ at different $\ve$ .
}\end{figure}

\begin{figure}
\caption{The  illustrative definition of quasilocality of interaction
in the lattice system. $\Ll$ and $\Gamma$ are any finite subsets of 
a lattice $\cal L$ .  The idea is that in case of large $\Gamma$
there is not observable influence on the configuration in $\Ll$ 
coming from from any  configuration $\s_{\Gamma^C}$ which is 
outside to the fixed configuration of $\s_\Gamma $.
}\end{figure}

\begin{figure}
\caption{ TCA with $(+)$ boundary added after reaching stabilisation. 
Distinct curves correspond to different values of stochastic 
perturbation $\ve$. With $\ve'$ we measure the extra magnetisation 
observed at the origin {\bf O} in stationary TCA.
}\end{figure}

\begin{figure}
\caption{Interaction between two TCA systems: $in$ and $out$ 
distinct from each other because of different stochastic parameters: 
$\ve_{in}$ and $\ve_{out}$, suitable.
a) $in$ initial state is $(+)$- phase, $\ve_{in}=0.86$, $out$ 
initial states are changed. Labels in the figure correspond to 
different values of $\ve_{out}$.
b) Propagation of self-created clusters of one phase from $in$ area 
into the $out$ configuration observed as random changes
in magnetisation of $out$ state.
}\end{figure}

\begin{figure}
\caption{ The relative entropy density between $\mu_+$ and $\mu_-$ 
stationary measures in the critical regime of TCA versus block size 
for lattices $L=60,\ 100,\ 20$.  
}\end{figure}

\begin{figure}
\caption{ The relative entropy density between $\mu_+$ and $\mu_-$ 
stationary measures  for large blocks.
}\end{figure}
\begin{figure}
\caption{Density of the relative entropy between stationary measures 
$\mu_-$ and $\nu$--- the measure for the system moved a little from
the critical point.
}\end{figure}

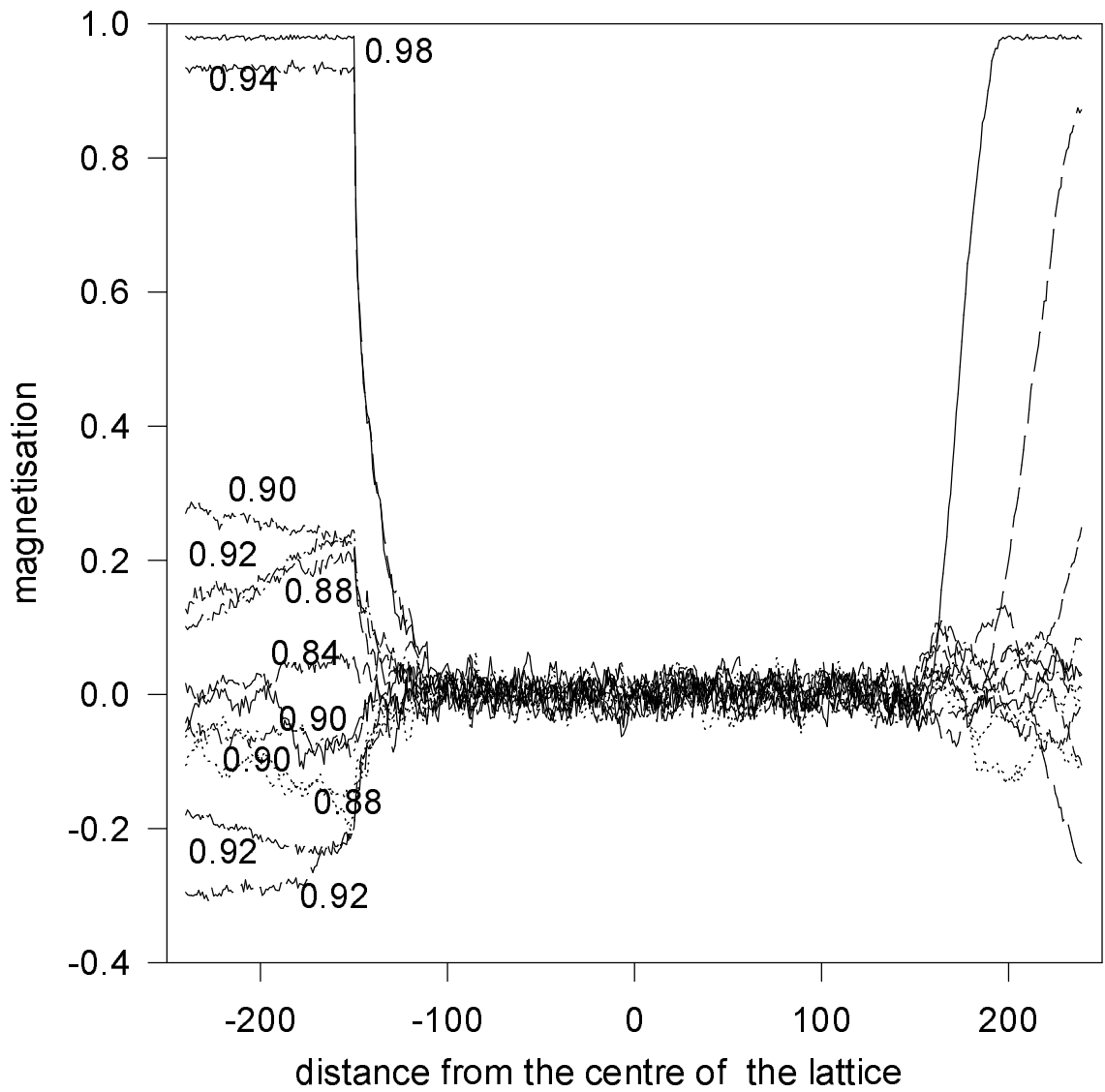
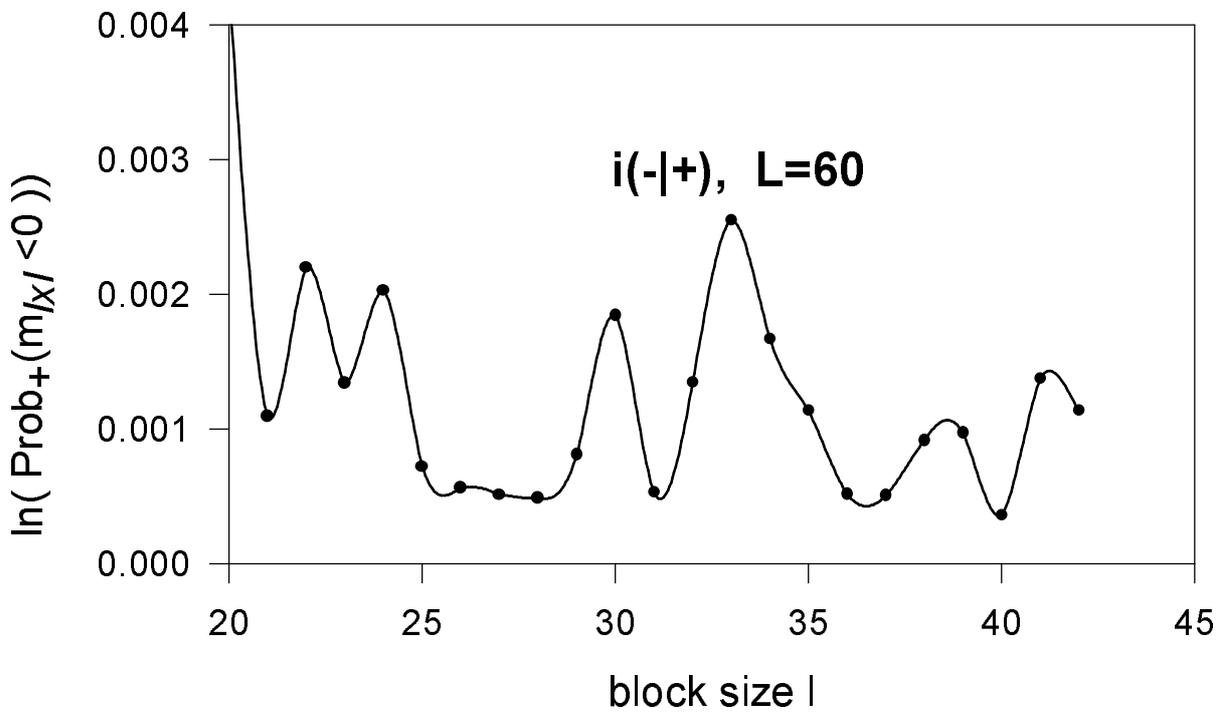
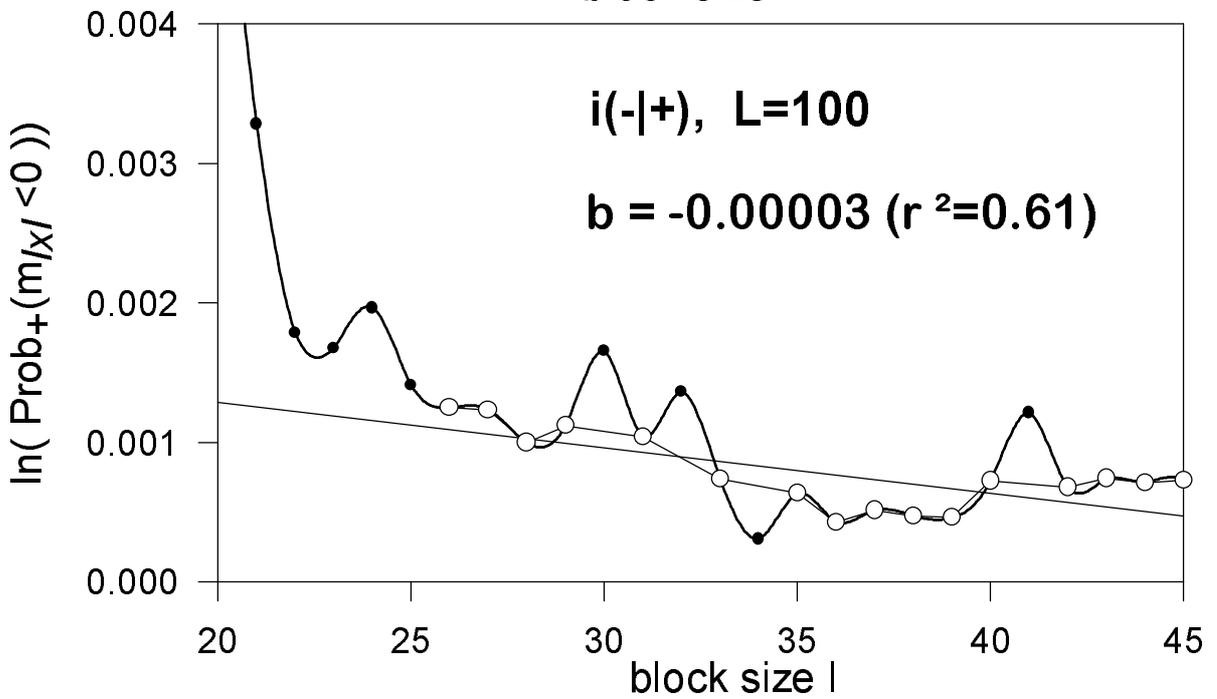
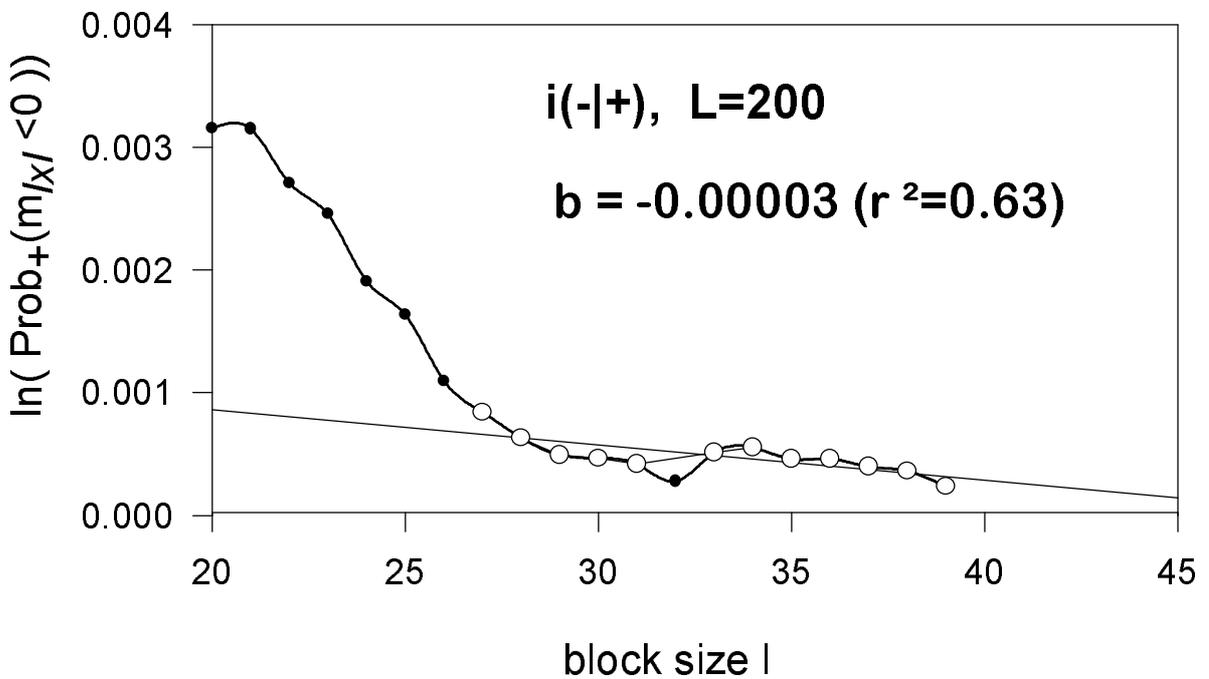
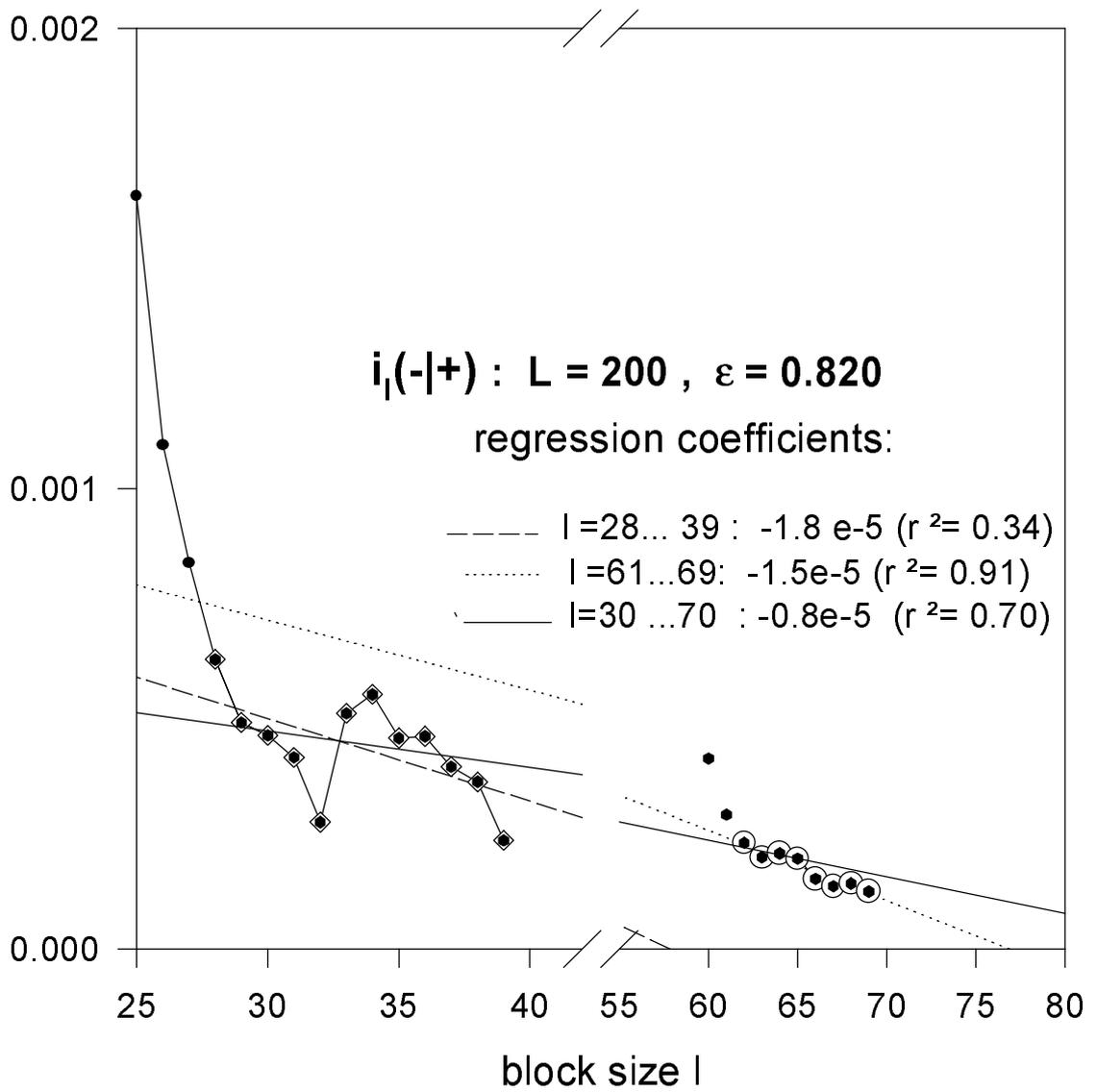
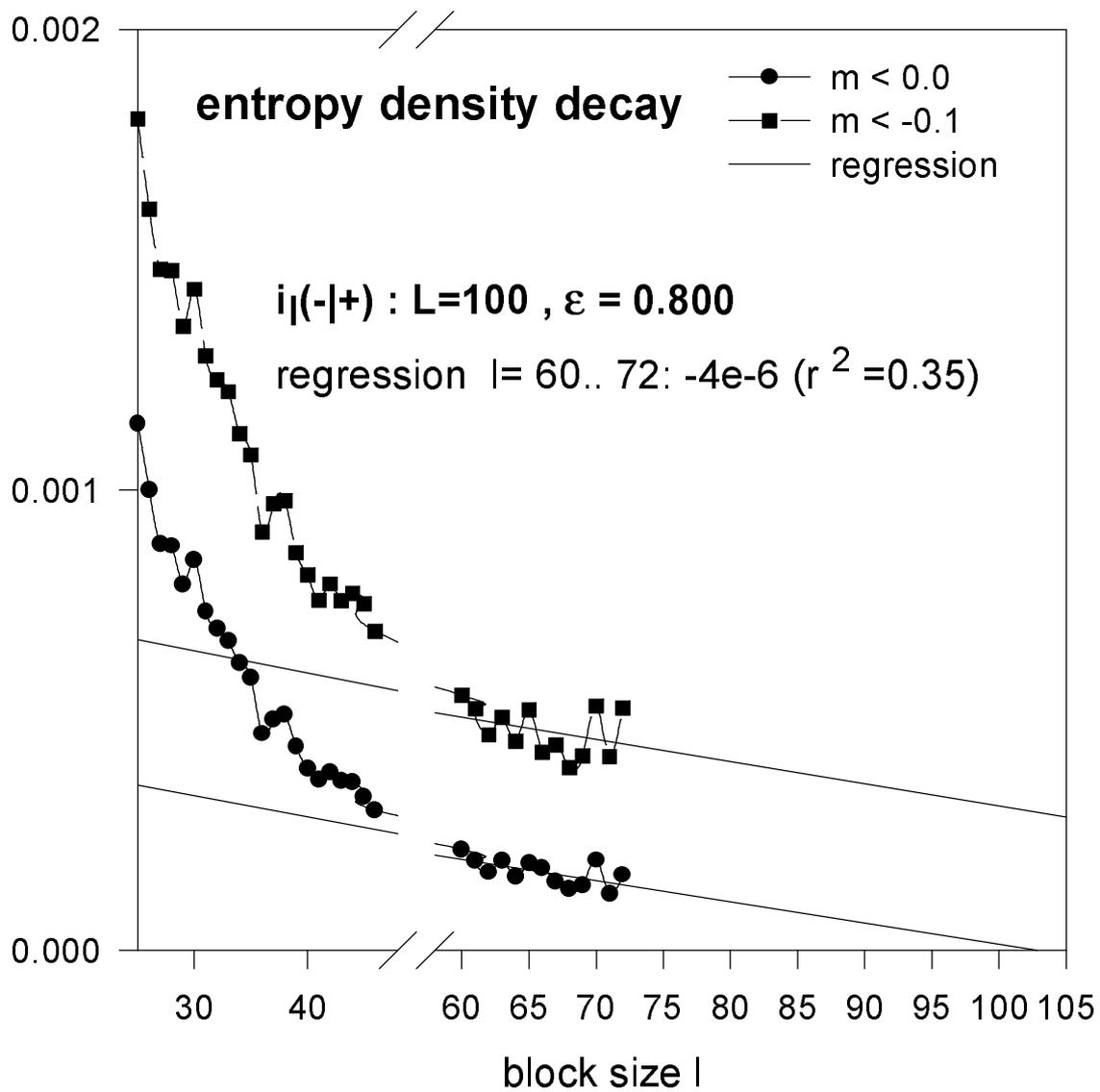
\end{document}